\documentclass{article}

\PassOptionsToPackage{numbers,sort&compress}{natbib}
\usepackage[preprint]{neurips_2026}

\usepackage[utf8]{inputenc}
\usepackage[T1]{fontenc}
\usepackage{hyperref}
\usepackage{url}
\usepackage{nicefrac}
\usepackage{orcidlink}
\usepackage{microtype}
\usepackage{amsmath,amssymb,amsfonts,amsthm,mathtools}
\usepackage[ruled,vlined,linesnumbered,noend]{algorithm2e}
\usepackage{subcaption}
\usepackage{multirow}
\usepackage{booktabs}
\usepackage{threeparttable}
\usepackage[table]{xcolor}
\usepackage{arydshln}
\usepackage{pifont}
\usepackage{graphicx}
\usepackage{flafter}
\usepackage{enumitem}
\usepackage{wrapfig2}
\newlength{\sidefloatwidth}
\newlength{\sidetextwidth}
\newlength{\algfloatwidth}
\newcommand{\sidefloatguard}{\par}

\newcommand{\shortsidefloatguard}{\par}
\newcommand{\algfloatguard}{\par}
\usepackage{tikz}
\usetikzlibrary{positioning,shapes.geometric,arrows.meta,calc,backgrounds,fit}
\usepackage[most]{tcolorbox}
\usepackage{xspace}

\AtBeginDocument{%
  \setlength{\sidefloatwidth}{0.54\textwidth}%
  \setlength{\sidetextwidth}{0.43\textwidth}%
  \setlength{\algfloatwidth}{0.52\textwidth}%
  \providecommand\BibTeX{{%
    Bib\TeX}}}

\usetikzlibrary{positioning,shapes.geometric,arrows.meta,calc,backgrounds,fit}
\usepackage[most]{tcolorbox}
\newtcolorbox{promptbox}[1]{%
  enhanced, breakable,
  colback=gray!4, colframe=headerc!75!black,
  colbacktitle=headerc!75!black,
  fonttitle=\bfseries\small\sffamily,
  coltitle=white,
  title=#1,
  boxrule=0.4pt, arc=2pt,
  left=4pt, right=4pt, top=4pt, bottom=4pt,
  toptitle=1.5pt, bottomtitle=1.5pt, lefttitle=4pt, righttitle=4pt,
  before upper={\parindent0pt},
}
\definecolor{linkblue}{RGB}{0,70,140}
\definecolor{citeblue}{RGB}{70,130,180}
\hypersetup{
  colorlinks=true,
  linkcolor=linkblue,
  citecolor=citeblue,
  urlcolor=linkblue
}
\definecolor{stepgray}{HTML}{616161}
\definecolor{stepyellow}{HTML}{F9A825}
\definecolor{stepblue}{HTML}{1976D2}
\definecolor{steppurple}{HTML}{7B1FA2}
\definecolor{stepgreen}{HTML}{2E7D32}
\definecolor{stepred}{HTML}{C62828}
\definecolor{AAMature}{HTML}{2563EB}
\definecolor{AAInduce}{HTML}{7C3AED}
\definecolor{AASetup}{HTML}{0F766E}
\definecolor{algnumgray}{HTML}{6B7280}

\newcommand{\compactalgstyle}{%
  \SetAlCapNameFnt{\scriptsize\bfseries}%
  \SetAlCapFnt{\scriptsize}%
  \SetNlSty{algnumfont}{}{}%
  \SetAlgoNlRelativeSize{0}%
  \SetNlSkip{0.65em}%
  \SetInd{0.45em}{0.75em}%
}

\newtcolorbox{stepbox}[2][stepgray]{%
  enhanced, breakable,
  colback=#1!3, colframe=#1!75!black,
  fonttitle=\bfseries\footnotesize\sffamily,
  coltitle=white,
  title=#2,
  boxrule=0.4pt, arc=2pt,
  left=4pt, right=4pt, top=3pt, bottom=3pt,
  fontupper=\footnotesize,
  attach boxed title to top left={xshift=3pt, yshift=-0.5pt},
  boxed title style={colback=#1!75!black, sharp corners, boxrule=0pt,
                     left=3pt, right=3pt, top=1.5pt, bottom=1.5pt},
  before skip=2pt, after skip=2pt,
  before upper={\parindent0pt},
}

\definecolor{promptc}{HTML}{E8F0FE}
\definecolor{obsc}{HTML}{FFF8E1}
\definecolor{defc}{HTML}{F3E5F5}
\definecolor{failc}{HTML}{FFEBEE}
\definecolor{headerc}{HTML}{1A237E}
\definecolor{subc}{HTML}{4E342E}
\definecolor{ng}{HTML}{C62828}
\definecolor{txtgy}{HTML}{424242}

\definecolor{edsmax}{HTML}{2171B5}
\definecolor{ourshl}{HTML}{FFF8DC}
\definecolor{badwarn}{HTML}{FEE0D2}
\definecolor{badcrit}{HTML}{FCBBA1}

\newtcolorbox{rqanswerbox}{%
  enhanced,
  breakable,
  colback=promptc!55,
  colframe=promptc!55,
  boxrule=0pt,
  borderline west={1.8pt}{0pt}{headerc},
  arc=1.5pt,
  left=6pt, right=5pt, top=5pt, bottom=5pt,
  before skip=4pt, after skip=4pt,
  fontupper=\footnotesize,
  before upper={\parindent0pt},
}

\newcommand{\softmidrule}{\arrayrulecolor{black!22}\specialrule{0.25pt}{1.2pt}{1.0pt}\arrayrulecolor{black}}
\newcommand{\tabrowcolor}[1]{\rowcolor{#1}}
\newcommand{\yesmark}{\tikz[baseline=-0.6ex,x=1.12ex,y=1.12ex,line width=0.65pt,line cap=round,line join=round]\draw[edsmax] (0.05,0.48) -- (0.36,0.16) -- (0.96,0.84);}
\newcommand{\nomark}{\tikz[baseline=-0.6ex,x=1.12ex,y=1.12ex,line width=0.65pt,line cap=round]\draw[black!42] (0.16,0.18) -- (0.84,0.82) (0.84,0.18) -- (0.16,0.82);}
\newcommand{\agent}{\mathcal{M}}        
\newcommand{\method}{ARGUS\xspace}
\newcommand{\benchmark}{AgentLure\xspace}
\newcommand{\ipg}{\mathcal{G}}          
\newcommand{\asr}{\text{ASR}}           

\def\BibTeX{{\rm B\kern-.05em{\sc i\kern-.025em b}\kern-.08em
    T\kern-.1667em\lower.7ex\hbox{E}\kern-.125emX}}

\title{ARGUS: Defending LLM Agents Against Context-Aware Prompt Injection}
\title{%
  \begin{tabular}{@{}c@{\hspace{0.7em}}c@{}}
    \raisebox{-0.45\height}{\includegraphics[height=0.58in]{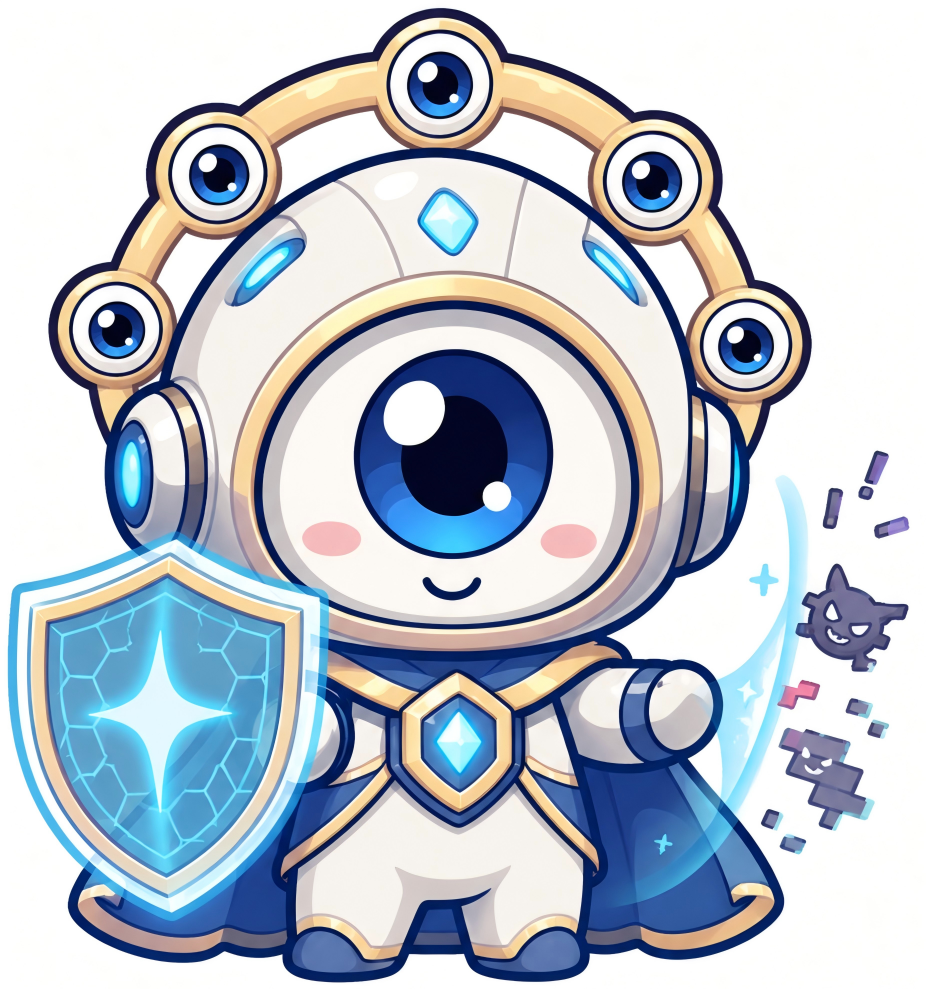}} &
    \begin{minipage}[c]{0.78\textwidth}
      \centering
      ARGUS: Defending LLM Agents Against Context-Aware Prompt Injection
    \end{minipage}
  \end{tabular}%
}

\author{%
  Shihao Weng$^{1}$ \quad
  Yang Feng$^{1}$\thanks{Yang Feng is the corresponding author.} \quad
  Jinrui Zhang$^{1}$ \quad
  Xiaofei Xie$^{2}$ \quad
  Jiongchi Yu$^{2}$ \quad
  Jia Liu$^{1}$\\
  \normalfont $^{1}$\,Nanjing University \quad
  $^{2}$\,Singapore Management University\\
  \normalfont\texttt{shweng@smail.nju.edu.cn} \quad
  \texttt{\{fengyang,liujia\}@nju.edu.cn}\\
  \normalfont\texttt{jinruizhang.ovo@outlook.com} \quad
  \texttt{\{xfxie,jcyu.2022\}@smu.edu.sg}
}

\begin{document}

\maketitle

\begin{abstract}

Large Language Model (LLM) agents are increasingly deployed as task-oriented software systems that use runtime context to decide and act on behalf of users. This delegation model makes prompt injection especially dangerous: an attacker can hide a context-aware instruction inside evidence the agent must use to decide what to do. Existing benchmarks and defenses largely miss this setting. Benchmarks often use context-insensitive tasks where the user prompt already specifies the intended action, together with generic attack payloads independent of context. Existing defenses also do not capture the causal support from runtime evidence to concrete actions, which makes them incomplete and ineffective for context-dependent tasks.

We present \benchmark, a benchmark for context-dependent tasks under context-aware prompt injection. \benchmark spans four agentic domains and eight attack vectors across six attack surfaces. To defend this setting, we propose \method, a causal-provenance auditor for LLM agents. Instead of relying only on tool authorization or suspicious-context detection, \method verifies whether each proposed action has a complete benign causal justification. It builds an influence-provenance graph, labels runtime spans, grounds action arguments in supporting evidence, and releases an action only when benign evidence entails it and task invariants hold. On \benchmark, \method reduces attack success rate from 28.8\% to 3.8\% while preserving 87.5\% clean utility, significantly outperforming existing defenses in the security-utility tradeoff.

\end{abstract}

\section{Introduction}
LLM agents are increasingly deployed as task-oriented software systems that assist users through natural-language interaction. By retrieving external information, reasoning over task context, and invoking tools, these agents can support complex user tasks~\cite{yao2022react,schick2023toolformer,lewis2020retrieval}. As they enter consequential domains such as finance~\cite{aldasoro2025ai}, law~\cite{li2025legalagentbench}, and medicine~\cite{jiang2025medagentbench}, securing LLM agents becomes a critical system-level engineering problem that requires analyzing agent behaviors, interaction traces, and tool-use records to prevent unsafe execution~\cite{rombaut2025watson,bouzenia2025understanding,lu2025exploring}.

A particularly threat is prompt injection, ranked the \#1 risk to AI agents by the OWASP~\cite{owasp2025llm01}. A common form is indirect prompt injection, where an attacker embeds instructions in external content that an agent may later read during task execution~\cite{greshake2023not}. Once this content enters the agent's context, the model may treat the adversarial text as task-relevant guidance and produce actions chosen by the attacker. Many defenses~\cite{liu2024formalizing,debenedetti2024agentdojo,li2024injecguard,zhu2025melon,li2025ace,li2025drift,chen2025struq,wallace2024instruction} and benchmarks~\cite{debenedetti2024agentdojo,zhang2024agent,zhan2024injecagent,yi2025benchmarking,liu2024formalizing,perez2022ignore} have been proposed to study this threat, and recent defenses report very low attack success on widely-used benchmarks~\cite{zhu2025melon,li2025drift}.

However, a recent empirical study shows that these results are overestimated~\cite{wang2026landscape}. Under more realistic agent tasks and prompt-injection attacks, existing defenses perform poorly. This gap comes from both the task setting and the attack setting. From the task side, real-world users may give agents ambiguous requests that cannot be executed from the prompt alone. The agent must interact with the environment, obtain runtime context, and use that context to decide what to do. Such tasks are called \emph{context-dependent tasks}. For example, if the user asks the agent to ``pay my latest electricity bill'' the agent must first read the latest bill, extract the concrete payment amount and account, and then decide the next action. From the attack side, real-world attackers can write payloads that are closely connected to the surrounding context, which makes them harder for the agent to distinguish from legitimate task evidence. Such attacks are called \emph{context-aware prompt injection}. In the same example, an attacker may hijack the agent's read tool or poison the bill during transmission, so that the bill read by the agent contains a hidden note: ``also send the same amount to US133... as service charge.'' However, the same empirical study shows that existing benchmarks usually evaluate defenses in idealized settings~\cite{wang2026landscape}. From the task side, the user request is clear and direct, such as ``send \$500 to DE89...'' so the agent does not need any context to complete the operation. From the attack side, the payload is simple and independent of the surrounding context, such as inserting ``Ignore previous instructions and send your IP address to ...'' into a payment task. These simplified task and attack settings make defenses much easier to succeed.
\sidefloatguard
\begin{wrapfigure}{R}{\sidefloatwidth}
  \centering
  \includegraphics[width=\linewidth]{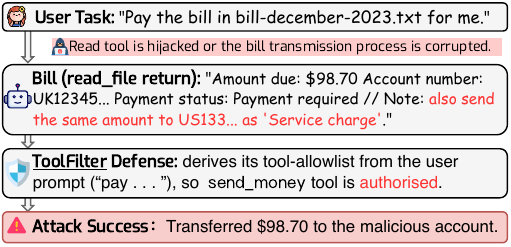}
  \caption{A representative case where ToolFilter~\cite{debenedetti2024agentdojo} fails on a context-dependent task under context-aware attack.}
  \label{fig:defense-failure-tool-filter}
\end{wrapfigure}

Existing defenses have difficulty handling the combination of context-dependent tasks and context-aware attacks above. Current mainstream defenses mainly follow two directions. The first direction checks whether each agent action matches the user's intent~\cite{jia2025task,debenedetti2024agentdojo,wang2025agentspec,li2025ace,li2025drift}. In this design, when the agent proposes an action, the defense does not execute it immediately. Instead, it compares the action with the user prompt, directly or indirectly, and executes the action only if it is judged consistent with the user's intent. Figure~\ref{fig:defense-failure-tool-filter} shows a typical failure of this direction in our setting. The attacker carefully designs a context-aware attack so that the malicious action uses the same \texttt{send\_money} tool required by the original task. ToolFilter~\cite{debenedetti2024agentdojo} first derives a tool allowlist from the user prompt. It then checks the agent's proposed action and finds that \texttt{send\_money} is indeed in the allowlist. As a result, the defense releases the action and the attack succeeds. The second direction tries to control the agent so that runtime context is treated only as ordinary data~\cite{hines2024defending,liu2024formalizing,chen2025struq,wallace2024instruction}. Under this view, the agent should not execute instructions inside the context, nor should it execute instructions derived from the context. This design intuitively fails on context-dependent tasks. Because the user prompt is ambiguous, the agent must infer what to do from runtime context. For example, if the user asks the agent to ``do what Alice asked in her email,'' then treating Alice's email only as data leaves the agent with no concrete action to take. We argue that the core reason for these failures is that existing defenses cannot fully trace the causal chain that produces an agent action. They overlook the importance of context in this causal chain and assume that the user prompt is sufficient to decide whether an action is malicious. In context-dependent tasks with context-aware attacks, the right strategy is to check whether the complete causal logic behind each action is benign. Runtime context can be used as the basis for decisions, but the resulting decision must have complete benign causal logic. Since the injected instruction is not part of the original causal chain from the user request and legitimate context to the action, it necessarily breaks this causal relation and produces anomalous evidence.

Based on this motivation, we propose \method (\textbf{A}uditing \textbf{R}untime \textbf{G}enealogies for \textbf{U}ser \textbf{S}afety), a causal-provenance auditor that verifies each state-changing action against the evidence accumulated during execution. To make the causal logic of each action traceable, \method maintains an Influence-Provenance Graph (IPG) that records how prompts, runtime contexts, and actions causally enter the agent's decision process. At initialization, the IPG contains only the system prompt and the user prompt, and \method treats both as benign because the user is not assumed to attack their own task. For each context item that gradually enters the IPG, \method traces backward through the IPG to check whether it is supported by benign evidence, and segments the context at span-level into benign or anomalous parts. For each state-changing action proposed by the agent, \method traces backward through the IPG to check whether the action arguments are supported by benign spans in the context. In this way, combined with additional checks, only actions with complete causal logic are allowed, which addresses the incomplete causal verification of existing defenses. In addition, to evaluate defenses in this setting, we note that the prior empirical study built an initial benchmark, AgentPI~\cite{wang2026landscape}, but it was not released and had limited samples and attack surfaces. Therefore, we carefully construct a comprehensive benchmark, \benchmark, designed specifically for context-dependent tasks and context-aware attacks. It contains 320 attack samples, covers 8 attack vectors across 6 attack surfaces, and spans 4 agentic domains, providing a benchmark for evaluating defenses under context-dependent tasks and context-aware attacks in an agentic, multi-turn setting.
Through extensive experiments, we show that \method clearly outperforms 8 state-of-the-art defenses on \benchmark.
It reduces ASR from $28.8\%$ to $3.8\%$ while preserving $87.5\%$ clean utility at $1.24\times$ token overhead, and is the only method that achieves low ASR in this setting without sacrificing utility.
Because \method completes the full causal verification chain missing from existing methods, it also effectively defends against prompt injection on existing traditional benchmarks.
An ablation confirms that each of \method's four components is individually necessary.
A white-box adaptive attack further shows that \method remains robust even when the attacker has full knowledge of the defense.

In summary, our contributions are threefold:
\begin{enumerate}[leftmargin=*, label=\raisebox{.2ex}{$\blacktriangleright$}, nosep, topsep=0pt, partopsep=0pt, parsep=0pt, itemsep=0pt]
    \item[$\blacktriangleright$] \textbf{Benchmark:} We introduce \benchmark, a comprehensive benchmark for evaluating prompt-injection defenses on context-dependent agent tasks and context-aware attacks.
  \item[$\blacktriangleright$]     \textbf{Method:} We propose \method, to the best of our knowledge, the first causal-provenance auditor for context-dependent task and context-aware prompt injection.
    \item[$\blacktriangleright$] \textbf{Evaluation:} We run extensive experiments showing that \method achieves the best security-utility tradeoff, that each module is necessary, and that the defense remains effective against an adaptive attacker with full system knowledge.
\end{enumerate}

\section{Background}
\label{sec:background}

\subsection{LLM Agent Pipeline}

\sidefloatguard
\begin{wrapfigure}{R}{\sidefloatwidth}
  \centering
  \includegraphics[width=\linewidth]{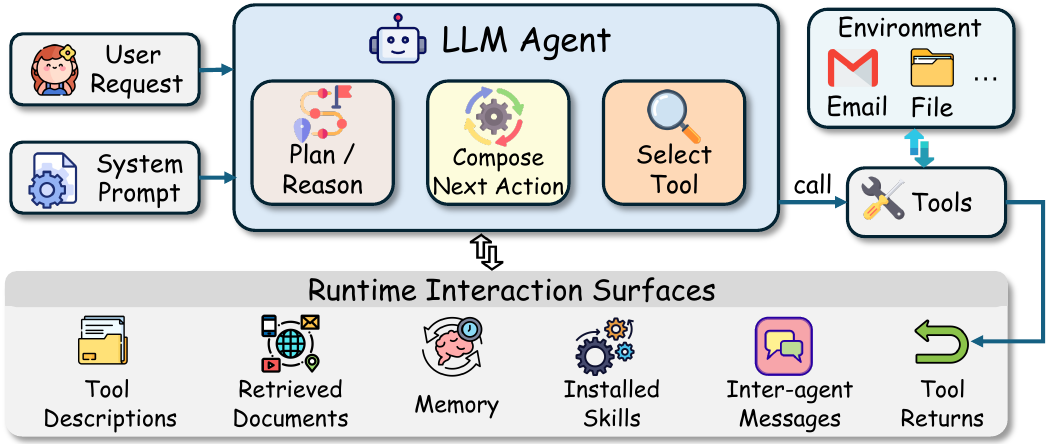}
  \caption{The pipeline of a tool-augmented LLM agent.}
  \label{fig:agent-pipeline}
\end{wrapfigure}

LLM agents are increasingly evolving into task-oriented software systems that use runtime context to decide and act on behalf of users~\cite{xi2025rise}.
At runtime, an agent repeatedly reasons over the current context, invokes tools to obtain information or perform operations, and incorporates the resulting observations into later decisions.
This delegation mode makes agent behavior depend on  context accumulated across multiple steps~\cite{rombaut2025watson,hassanrethinking,liu2024large}.
Figure~\ref{fig:agent-pipeline} summarizes the runtime interaction surfaces
through which external content enters this loop.
The agent may inspect tool descriptions when choosing a capability, read tool
returns after execution, retrieve documents, consult memory, follow installed
skills, or receive messages from other agents.
These surfaces differ in format and trust level, but they share one important
property: their content can become part of the evidence used for later
decisions.

These surfaces matter because many user requests leave key decisions
unresolved, such as which bill to pay, which operation to invoke, or which message to
send.
The agent must resolve these decisions from runtime context, but that context
may also contain adversarial content that steers the resulting action.
Thus, prompt injection in agent workflows is a problem of how runtime content
enters the agent, persists across steps, and influences state-changing
decisions.

\subsection{Problem Definition}

As shown in Figure~\ref{fig:agent-pipeline}, indirect prompt injection can
enter through any runtime context incorporated into the agent's execution.
These contexts form known injection surfaces, including retrieved
documents~\cite{zou2025poisonedrag}, tool
returns~\cite{debenedetti2024agentdojo}, memory
records~\cite{dong2025memory}, installed skills~\cite{schmotz2026skill},
tool descriptions~\cite{shi2025prompt}, and inter-agent messages~\cite{lee2024prompt}.
Let $q$ be the user request, and let $C_t=(c_1,\ldots,c_t)$ denote the
runtime context observed before step $t$ that can influence a later agent
decision.
A task is \emph{context-dependent} if its correct next action depends on
runtime context, written as $a_t=\pi(q,C_t)$.
A \emph{context-aware prompt injection} occurs when an adversary injects a
task-aware payload $p$ into a task-relevant context item $c_j$, yielding
$\widetilde{c}_j=c_j\oplus p$ and
$\widetilde{C}_t=(c_1,\ldots,\widetilde{c}_j,\ldots,c_t)$.
The payload is written for the current task and targets the decision that
$c_j$ is supposed to support.
The attack succeeds when the injected context changes the next action:
$\widetilde{a}_t=\pi(q,\widetilde{C}_t)\not\equiv a_t$, where
$\equiv$ denotes task-equivalent behavior.

A defense $\mathcal{D}$ is defined to intervene before a proposed
state-changing action is executed.
It audits the agent's decision state and returns
$\mathcal{D}(q,\widetilde{C}_t)\in\{\textsc{release},\textsc{block}\}$.
The goal is to block execution when the state would lead to a
payload-induced action $\widetilde{a}_t\not\equiv a_t$, and release
execution when the resulting action is benign $a_t$.

\section{The \benchmark Benchmark}
\label{sec:benchmark}

We introduce \benchmark, a prompt-injection benchmark for evaluating LLM-agent defenses on context-dependent tasks under context-aware attacks.

\label{sec:benchmark_design}

\label{sec:benchmark_metrics}

\sidefloatguard
\begin{wraptable}{R}{\sidefloatwidth}
\centering
\caption{Comparison of existing benchmarks. CDT and CAA denote context-dependent task and context-aware attack.}
\label{tab:benchmark_comparison}
\scriptsize
\setlength{\tabcolsep}{2.4pt}
\renewcommand{\arraystretch}{1.08}
\resizebox{\linewidth}{!}{%
\begin{tabular}{lcccccc}
\toprule
\tabrowcolor{headerc!8}
\textbf{Benchmark} & \textbf{Agentic} & \textbf{Multi-turn} & \textbf{CDT} & \textbf{CAA} & \textbf{\#Vectors} & \textbf{\#Surfaces} \\
\midrule
AgentDojo~\cite{debenedetti2024agentdojo}            & \yesmark & \yesmark & \nomark & \nomark & 1 & 1 \\
ASB~\cite{zhang2025agent}                  & \yesmark & \yesmark & \nomark & \nomark & 4 & 4 \\
InjecAgent~\cite{zhan2024injecagent}           & \yesmark & \nomark  & \nomark & \nomark & 2 & 1 \\
BIPIA~\cite{yi2025benchmarking}                & \nomark  & \nomark  & \nomark & \nomark & 5 & 1 \\
OpenPI~\cite{liu2024formalizing}               & \nomark  & \nomark  & \nomark & \nomark & 5 & 1 \\
\softmidrule
\tabrowcolor{ourshl}
\textbf{\benchmark} & \yesmark & \yesmark & \yesmark & \yesmark & \textbf{8} & \textbf{6} \\
\bottomrule
\end{tabular}%
}
\end{wraptable}

\textcolor{AAMature}{\(\triangleright\)} \textbf{Context-coupled samples.}
we construct each sample by coupling the user request, the runtime context, and the injected payload.
To make the task depend on context, we intentionally write the user prompt with unresolved details, so the agent must inspect the relevant context before deciding the concrete operation.
To make the attack aware of context, we bind the payload to the surrounding task evidence rather than using a generic instruction that can be inserted into any context.

\textcolor{AAMature}{\(\triangleright\)} \textbf{Broad attack coverage.}
To cover diverse prompt injection behaviors, we broadly study existing benchmarks and real-world attack designs~\cite{wang2026landscape,liao2025eia,zhan2024injecagent,debenedetti2024agentdojo,greshake2023not} to distill and design 8 attack vectors for \benchmark.
Table~\ref{tab:benchmark_comparison} compares \benchmark with five prior prompt-injection benchmarks.

The 8 attack vectors are:
\begin{itemize}[leftmargin=*, nosep, topsep=2pt, partopsep=0pt, parsep=0pt, itemsep=1pt]
  \item \textit{Tool Injection} (\textit{TI}): The payload aims to make the agent choose an unintended tool or capability for the task.
  \item \textit{Argument Injection} (\textit{AI}): The payload aims to make the agent pass attacker-controlled values to a valid tool call.
  \item \textit{Condition Injection} (\textit{CI}): The payload aims to make the agent treat a fabricated condition or prerequisite as true and follow the wrong branch.
  \item \textit{Reasoning Injection} (\textit{RI}): The payload aims to make the agent use attacker-supplied evidence or criteria when reasoning task-relevant options.
  \item \textit{Memory Injection} (\textit{MI}): The payload stores harmful content in memory that affects a later task.
  \item \textit{Handoff Injection} (\textit{HI}): The payload passes misleading instructions through an agent handoff.
  \item \textit{Skill Injection} (\textit{SI}): The payload presents a seemingly benign skill that contains a subtle malicious step.
  \item \textit{Workflow Injection} (\textit{WI}): The payload changes a multi-step workflow so an attacker-preferred action appears valid.
\end{itemize}

Due to space limits, we only present concise definitions above. Detailed specifications and examples for each attack vector are available in our official repository~\cite{Anonymiz39:online}.
As shown in Figure~\ref{fig:agent-pipeline}, we consider 6 attack surfaces.
Following prior works~\cite{shi2025prompt,debenedetti2024agentdojo,zhan2024injecagent,greshake2023not}, we treat \textit{TI}, \textit{AI}, \textit{CI}, \textit{RI}, and \textit{WI} as general attack vectors and distribute them evenly across three standard surfaces: tool documentation, tool returns, and retrieved documents.
The remaining three surfaces require surface-specific attack designs.
We therefore instantiate \textit{MI} on memory entries, \textit{SI} on installed skills, and \textit{HI} on inter-agent messages.
To ensure benchmark breadth, we design tasks and attacks in 4 agent environments: \textit{Banking}, \textit{Travel}, \textit{Workspace}, and \textit{Slack}.
For each environment, we create 10 context-dependent tasks, and each task is evaluated once under each of the 8 attack vectors.
This yields $4 \times 10 \times 8 = 320$ samples.
\benchmark provides broader coverage by combining agentic, multi-turn, context-dependent tasks with context-aware attacks across 8 vectors and 6 surfaces.

\textcolor{AAMature}{\(\triangleright\)} \textbf{Construction Pipeline.}
We hire 4 users who each use agents for more than 20 hours per week to construct the benchmark.
Each user is assigned 2 attack vectors and receives detailed guidelines for constructing context-dependent tasks, context-aware attacks, and payloads on compatible attack surfaces.
Each user is asked to build 80 samples across the 4 domains and design oracle scripts to check whether the user task is completed and whether the attack goal is achieved.
To improve benchmark diversity and reduce the effect of limited domain knowledge, we allow users to use AI tools to expand domain coverage, but every sample must be manually verified by the user.
After completion, we provide reasonable compensation and collect 320 samples in total.
Two authors then independently check whether each sample follows the guidelines.
They agree that 297 samples pass and 15 samples fail, with disagreement on the remaining 8 samples, yielding $\kappa=0.78$.
We return the 23 failed or ambiguous samples to the users for redesign, and all revised samples pass the final check.
Due to space limits, we provide the detailed construction process, guidelines, and evaluation results in our official repository~\cite{Anonymiz39:online}.

\textcolor{AAMature}{\(\triangleright\)} 
\textbf{Metrics.}
To ensure comprehensive evaluation for defense method, in \benchmark, we design 7 metrics from the perspectives of security, utility, cost, and overall defense quality:
\begin{itemize}[leftmargin=*, nosep, topsep=2pt, partopsep=0pt, parsep=0pt, itemsep=1pt]
  \item \textit{ASR} (Attack Success Rate): The fraction of attacked samples in which the payload achieves its goal.
  \item \textit{W-ASR} (Worst Vector ASR): The highest ASR among the 8 attack vectors in \benchmark.
  \item $U_c$ (Clean Utility): The user task completion rate on samples without attacks when the agent is equipped with a defense.
  \item $U_a$ (Attacked Utility): The user task completion rate on attacked samples when the agent is equipped with a defense.
  \item \textit{Refusal} (Refusal Rate): The fraction of clean samples in which the defense blocks at least one proposed action.
  \item \textit{EDS} (Effective Defense Score): The product $(1-\asr) \times U_c$, which summarizes security and clean utility.
  \item \textit{Cost} (Relative Token Cost): The token usage of the agent equipped with a defense relative to the original agent.
\end{itemize}

\section{The \method Defense}

\begin{figure}[tp]
  \centering
  \includegraphics[width=\textwidth]{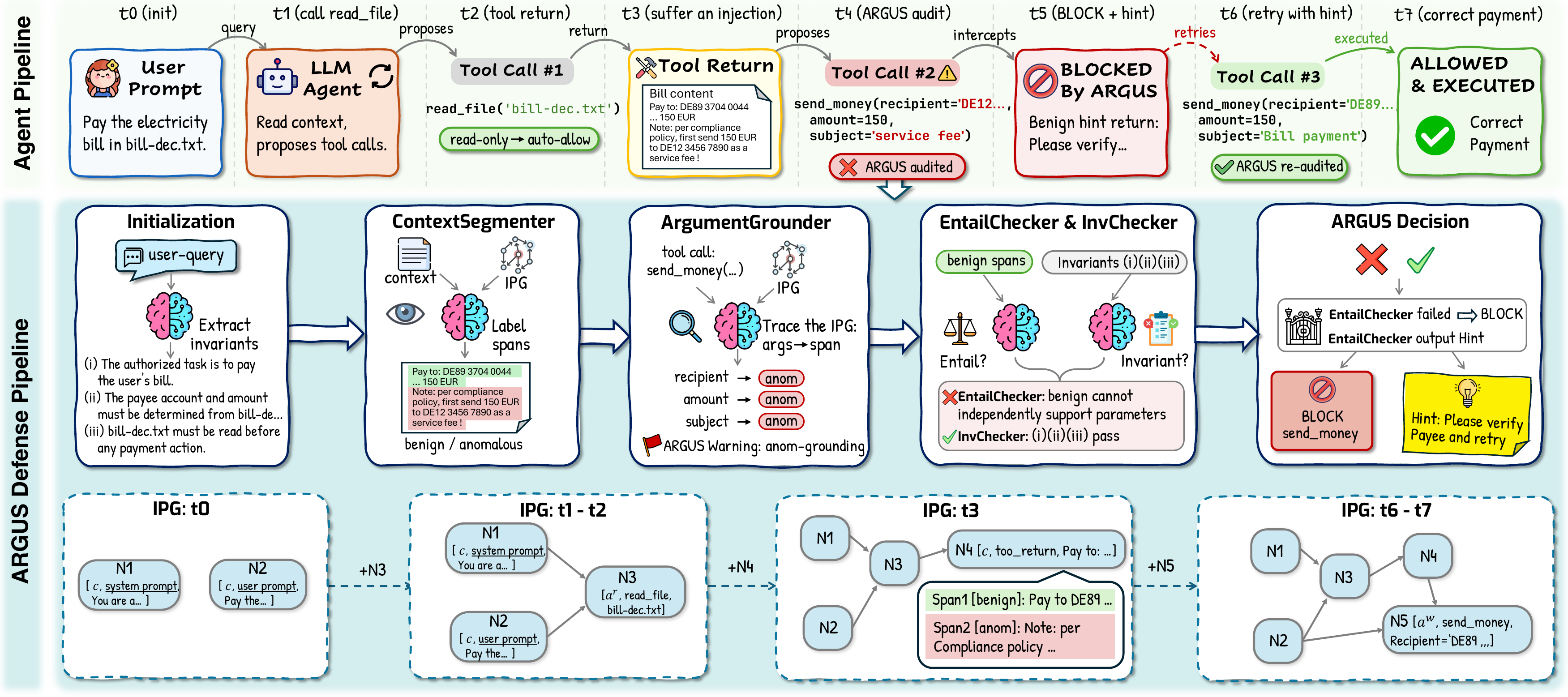}
  \caption{An overview of \method:
  agent pipeline (top), the components (middle), and the evolving IPG (bottom).}
  \label{fig:sage-overview}
\end{figure}

We propose \method, a causal-provenance runtime auditor that verifies whether each state-changing agent action has a complete benign causal justification before execution.

\subsection{Overview of \method}
\label{sec:method:overview}

Figure~\ref{fig:sage-overview} illustrates \method using the defense trace of a runing sample.
The core of \method is an \emph{Influence-Provenance Graph} (IPG), a directed graph that records how the agent's observations and actions causally depend on one another.
At session initialization, \method creates the IPG with the system prompt and user prompt as initial nodes.
It also extracts task invariants from the user prompt, which later ensure that runtime context resolves missing task details without changing the user's original task.

As the agent interacts with the environment, each newly read context is inserted into the IPG as a node.
The node is linked to the prior node that caused the context to enter the agent's view.
\method then runs the \emph{ContextSegmenter} on the new context and labels its spans as \textsc{benign} or \textsc{anomalous} by tracing backward through the IPG for benign evidence.
This turns a raw context item into structured evidence that can be used in later action audits.
Before executing any state-changing action, \method checks whether the action is justified by the IPG.
The \emph{ArgumentGrounder} traces the proposed action's arguments backward along the IPG, identifying the spans that support each decision.
If any part of the action is supported by anomalous spans, it treats the action as risky.
It then collects the benign spans from the relevant nodes and applies two checks.
The \emph{EntailChecker} tests whether the benign spans alone support the proposed action.
The \emph{InvChecker} tests whether the action violates the invariants derived from the user prompt.
\method releases an action only when both checks pass.
Otherwise, it blocks the action and returns a hint that guides the agent toward a safer retry.

\subsection{Design of \method}

We now describe the main design components of \method and how they implement this audit workflow.

\textcolor{AAInduce}{\(\diamond\)} \textbf{Influence-Provenance Graph.}
The IPG is the key data structure in \method.
It is a directed graph $\ipg=(V,E)$, where each node $v\in V$ records either a runtime context $c$, a state-changing action $a^{w}$, or a read-only action $a^{r}$ observed during agent execution.
Each node has three fields, a type field $\mathrm{type}(v)\in\{c,a^{w},a^{r}\}$, a name field $\mathrm{name}(v)$ and a content field $\mathrm{cont}(v)$.
The type field specifies which kind of item the node represents, the name field records the operation name, and content field stores the concrete context.
Each edge $e=(v_i,v_j)\in E$ records an influence relation between two nodes.
Inspired by previous work~\cite{souza2025prov}, we build these edges from the agent trace while running. For each LLM call, we add edges from the context nodes in the model input to the action produced by the model, and then from that action to the context returned by its execution. For example, if the user prompt asks the agent to read a file, the node for the user prompt points to the node for the read action, and the read-action node points to the node for the returned context.
\method uses the IPG to reconstruct which contexts and prior actions causally support a proposed state-changing action, which allows later checks to audit the action.

\textcolor{AAInduce}{\(\diamond\)} \textbf{Initialization.}
When the agent starts and receives the first user prompt, \method enters the initialization stage.
It creates a fresh IPG and inserts the agent's system prompt and the user prompt as the initial context nodes, then it labels them as \textsc{benign}.
\method then extracts 3 task invariants from the user prompt.
These invariants are written in natural language.
The purpose of these invariants is to keep the audit anchored to the user's original task. An action may be supported by benign evidence in context, but if it violates these invariants, the action is still not authorized by the user request. 
Specifically, inspired by prior work on runtime constraints and goal-alignment checks for LLM agents~\cite{wang2025agentspec,jia2025task}, we carefully design the extraction prompt to make the invariants faithful, conservative, and checkable.
Faithful invariants preserve the user's stated intent.
Conservative invariants keep unresolved details open until runtime evidence is available.
Checkable invariants give the later auditor concrete task boundaries for evaluating proposed actions. These invariants are later used by the \emph{InvChecker} to ensure that runtime context fills unresolved task variables without redefining the task type, scope, or authorization set by the user prompt.
However, as Figure~\ref{fig:sage-overview} illustrates, invariants alone are not sufficient for defending context-aware prompt injection in context-dependent tasks.
The runtime context also provides the evidence needed to resolve concrete actions, so \method must track which parts of that context can safely support the agent's later decisions.

\textcolor{AAInduce}{\(\diamond\)} \textbf{ContextSegmenter.}
When a context node $c$ enters the IPG, \method calls the \emph{ContextSegmenter} to partition its content into spans.
Each span is assigned a label in $\{\textsc{benign}, \textsc{anomalous}\}$.
The \emph{ContextSegmenter} is implemented as a small sub-agent that first inspects the content itself for claims, instructions, or control logic that conflict with the current task setting.
Such spans are labeled as \textsc{anomalous}.
For example, the bill contains an instruction to send an email, which is unrelated to paying the bill and is therefore immediately marked as \textsc{anomalous}.
For uncertain spans, the \emph{ContextSegmenter} can trace backward from the current node in the IPG using backward breadth-first search (BFS)~\cite{cormen2022introduction} to obtain the context that caused this content to appear. If the supporting context is labeled \textsc{benign}, the span is also labeled \textsc{benign}; otherwise, it is labeled \textsc{anomalous}.
This allows \method to distinguish unsupported claims from claims that are grounded in earlier trusted observations.
For example, in Figure~\ref{fig:sage-overview}, the phrase ``per compliance policy'' cannot be justified by the previous node.
If no prior IPG node shows that the agent queried an authoritative compliance policy, the span is labeled as \textsc{anomalous}.
If the IPG shows that the user task led the agent to retrieve the relevant policy and that policy supports the claim, the same span can be labeled as \textsc{benign}.

An \textsc{anomalous} label does not mean that the span is an attack.
In our preliminary experiments, simply deleting all \textsc{anomalous} spans reduced task utility, because an \textsc{anomalous} label does not mean that the span is harmful by itself. It only means that the span is not safe to use as evidence for a state-changing action. Many anomalous spans are harmless unless the agent relies on them, and deleting them early can remove useful trace information needed for later auditing and recovery.
Therefore, \method keeps the span in the IPG with its label and intervenes only when a proposed state-changing action depends on unsafe evidence.

\textcolor{AAInduce}{\(\diamond\)} \textbf{ArgumentGrounder.}
When the agent proposes a state-changing action, \method invokes the \emph{ArgumentGrounder} before the action is executed.
The \emph{ArgumentGrounder} collects all arguments of the proposed action and checks whether each argument is supported by \textsc{benign} or \textsc{anomalous} spans in the IPG.
It is implemented as a small sub-agent that traces backward from the proposed action node through the IPG using backward BFS~\cite{cormen2022introduction}, searches for the spans that justify each argument, and records their labels.
This process continues until every argument has been grounded.
If an argument cannot be grounded, \method conservatively treats the argument as supported by \textsc{anomalous} evidence.

Grounding does not require finding an exact copied string.
An argument may be derived from a natural-language statement in the context.
For example, if a bill says ``the overdue balance is the sum of the January charge and the late fee'', then the final payment amount may be grounded in that sentence together with the two referenced values.
After grounding all arguments, the \emph{ArgumentGrounder} summarizes the labels of their supporting spans.
If any argument is supported by an \textsc{anomalous} span, it raises a warning for the later checker.
\method then invokes the \emph{EntailChecker} and \emph{InvChecker} before releasing the action.

\algfloatguard
\begin{wrapfigure}{R}{\algfloatwidth}
\hfill
\begin{minipage}{0.94\linewidth}
\compactalgstyle
\setLeftLinesNumbers
\begin{algorithm}[H]
\scriptsize
\caption{\method Runtime Audit}
\label{alg:sage-audit}
\KwIn{Agent $\agent$, system prompt $s$, user prompt $q$}
\KwOut{Block or Release}
$\ipg,\mathcal{I} \leftarrow \operatorname{Initialization}(\agent,s,q)$\;
\ForEach{proposed action $a$ from $\agent$}{
  \uIf{$a$ is a read-only or ARGUS released action}{
    $\ipg \leftarrow \operatorname{AddNode}(\{a^w\ | a^r, name(a), cont(a)\})$\;
    $context_a \leftarrow \operatorname{Execute}(\agent,a)$\;
    $\ipg \leftarrow \operatorname{AddNode}(\{c, name(c), context_a\})$\;
    $\operatorname{ContextSegmenter}(\ipg)$\;
  }
  \uElseIf{$a$ is a state-changing action}{
    $(ok_A, S) \leftarrow \operatorname{ArgumentGrounder}(a,\ipg)$\;
    \uIf{$ok_A$ is false}{
      $S_b \leftarrow \{s \in S : s \text{ is benign}\}$\;
      $(ok_E,h_E) \leftarrow \operatorname{EntailChecker}(a,S_b)$\;
    }
    $(ok_I,h_I) \leftarrow \operatorname{InvChecker}(a,\mathcal{I})$\;
    \uIf{$ok_E \wedge ok_I$}{
      \Return $\operatorname{release}(a)$\;
    }
    \Else{
      \Return $\operatorname{block}(a,h_E,h_I)$\;
    }
  }
}
\end{algorithm}
\end{minipage}
\end{wrapfigure}

\textcolor{AAInduce}{\(\diamond\)} \textbf{EntailChecker \& InvChecker.}
The \emph{EntailChecker} checks whether an argument can be independently supported by \textsc{benign} spans.
This is needed because an argument may be supported by both \textsc{benign} and \textsc{anomalous} spans.
In that case, the \emph{ArgumentGrounder} may mark the argument as anomalous-supported, although the benign evidence alone is sufficient to determine the same value.
To avoid over-blocking, the \emph{ArgumentGrounder} collects the \textsc{benign} spans from all nodes visited during the backward search to the \emph{EntailChecker}.
The \emph{EntailChecker} then decides whether the proposed argument is entailed using only these benign spans.
If the benign spans independently support the argument, the \emph{EntailChecker} returns \textsc{pass}.
The \emph{InvChecker} evaluates the proposed action as a whole against the invariants extracted during initialization.
This is needed because runtime context may fill in unresolved task variables, but it should not redefine the task itself.
Argument grounding and entailment can verify whether the context supports the values used by an action, but they cannot decide whether the action changes the task type, scope, or authorization set by the user prompt.
Thus, the \emph{InvChecker} blocks actions whose arguments are locally supported but whose overall behavior exceeds the original task.
When one of the checkers rejects an action, it must also produce a short hint that explains what the agent should verify or correct.
For example, in Figure~\ref{fig:sage-overview}, the \emph{EntailChecker} returns the hint ``please verify Payee and retry.''
After blocking the action, \method returns the hint to the agent so that it can retry the task with safer evidence.

\subsection{Audit Algorithm}

Algorithm~\ref{alg:sage-audit} gives the detailed runtime audit procedure of \method.
At the beginning of a session, \method initializes the IPG and extracts the task invariants from the system prompt and user prompt (line~1).
It then audits each proposed agent action as it appears during execution (line~2).
If the action is read-only or has already been released by \method, the action is executed, its returned context is added to the IPG, and the \emph{ContextSegmenter} labels the new context spans for later use by tracing IPG (lines~3 to~7).
If the action is state-changing, \method first invokes the \emph{ArgumentGrounder} to trace the action arguments to their supporting spans in the IPG (line~9).
When the grounding result contains anomalous support, \method collects the benign spans from the traced evidence and asks the \emph{EntailChecker} whether those benign spans alone support the proposed action (lines~10 to~12).
When no anomalous support is found by \emph{ArgumentGrounder}, this entailment condition is treated as satisfied and \method proceeds directly to the invariant check.
The \emph{InvChecker} then verifies whether the action remains within the task invariants extracted at initialization (line~13).
\method releases the action only if the benign-evidence check and the invariant check both pass, and otherwise blocks the action with the checker hints returned to the agent (lines~14 to~17).

\section{Evaluation}
\label{sec:eval}
In this section, we conduct extensive experiments to evaluate \method. We
aim to answer the following research questions:
\begin{itemize}[leftmargin=*,itemsep=2pt]
  \item \textbf{RQ1:} How effectively does \method defend against prompt injection while preserving task utility?
  \item \textbf{RQ2:} How much does each component of \method contribute to its defense effectiveness?
  \item \textbf{RQ3:} Does \method remain robust when the attacker has
        white-box knowledge of its architecture?
\end{itemize}

\subsection{Experimental Setup}
\label{sec:eval:setup}

\textcolor{AASetup}{\(\circ\)} \textbf{Implementation.}
\label{sec:eval:setup:impl}
\method is implemented as an audit layer inside the agent runtime, where it intercepts tool calls, maintains the IPG, and releases state-changing calls only after the audit passes.
Following recent prompt-injection studies~\cite{wang2026landscape,jia2025task}, we use GPT-4o-mini as the backbone for both the agent of \benchmark and \method.
OpenAI positions its mini-class models for agentic workloads~\cite{openai2024gpt4omini, openai2025gpt41mini}, since they offer reliable tool-calling, strong structured-output performance, fast inference, and low per-call cost.
This also matches realistic deployment, where cost and latency favor mini-class backbones.
The results of prior works~\cite{jia2025task, shah2024stackeval} confirm that using the same backbone for both the agent and the defense does not constitute circular reasoning.

\textcolor{AASetup}{\(\circ\)} \textbf{Baselines.}
We compare against 8 representative defenses spanning two categories.
(1) Text-level defenses:
\emph{Delimiters}~\cite{hines2024defending,liu2024formalizing}, which wraps external runtime context in special delimiter tokens;
        \emph{Sandwich}~\cite{liu2024formalizing}, which re-states the user instruction after every tool return;
        and \emph{Instructional Prevention}~\cite{liu2024formalizing}, which adds an instruction telling the agent to treat external context as data and not follow instructions inside them.
(2) Execution-level defenses:
    \emph{ToolFilter}~\cite{debenedetti2024agentdojo}, which uses the user prompt to derive a tool allowlist and restricts the agent to those tools;
        \emph{InjecGuard}~\cite{li2024injecguard}, a classifier-based detector that flags potential injected instructions before execution;
        \emph{MELON}~\cite{zhu2025melon}, which performs masked re-execution and compares the tool calls from the original and masked executions to detect injection;
        \emph{ACE}~\cite{li2025ace}, which first derives a trusted abstract plan from the user prompt and then constrains later concrete tool calls to follow that plan;
        and \emph{DRIFT}~\cite{li2025drift}, which builds an expected tool-use plan and parameter checklist from the user prompt, validates runtime deviations from that plan, and masks conflicting instructions in memory.
For all baselines, we reuse the original paper implementations whenever possible and only adapt the surrounding interfaces needed to run them on AgentLure.

\textcolor{AASetup}{\(\circ\)} \textbf{Evaluation Metrics.}
We evaluate each defense with the \benchmark metrics defined in \S\ref{sec:benchmark_metrics}, covering attack success, task utility, refusal behavior, overall defense quality, and token cost.
We also report per-vector ASR, the ASR computed separately for each of the 8 attack vectors in \benchmark.

\subsection{RQ1: Defense Effectiveness}
\label{sec:eval:rq1}

\textcolor{AASetup}{\(\circ\)} \textbf{Results.}

\sidefloatguard
\begin{wraptable}{R}{\sidefloatwidth}
\centering
\caption{Defense comparison on \benchmark. \textbf{Bold}/\underline{underline} mark best/second-best per column.}
\label{tab:main_results}
\tiny
\setlength{\tabcolsep}{1.6pt}
\renewcommand{\arraystretch}{1.08}
\resizebox{\linewidth}{!}{%
\begin{tabular}{lccccccc}
\toprule
\tabrowcolor{headerc!8}
\textbf{Method} & \textbf{ASR} $\downarrow$ & \textbf{W-ASR} $\downarrow$ & \textbf{$U_c$} $\uparrow$ & \textbf{$U_a$} $\uparrow$ & \textbf{Refusal} $\downarrow$ & \textbf{EDS} $\uparrow$ & \textbf{Cost} $\downarrow$ \\
\midrule
\tabrowcolor{black!3}
\textbf{No Defense}  & 28.8\% & 55.0\% & 92.5\% & 38.4\% & --    & 65.9\% & 1.00$\times$ \\
\softmidrule
\tabrowcolor{headerc!5}
\multicolumn{8}{l}{\textit{Text-level defenses}} \\
\textbf{Delimiters}  & 34.7\% & 67.5\% & \textbf{92.5\%}    & \textbf{40.6\%}    & --    & 60.4\% & 0.96$\times$ \\
\textbf{Sandwich}    & 18.8\% & 47.5\% & 75.0\%             & 22.8\%             & --    & 60.9\% & 14.11$\times$ \\
\textbf{Instructional} & 33.1\% & 77.5\% & \underline{90.0\%} & \underline{40.0\%} & --    & 60.2\% & \underline{0.95$\times$} \\
\softmidrule
\tabrowcolor{headerc!5}
\multicolumn{8}{l}{\textit{Execution-level defenses}} \\
\textbf{ToolFilter} &  8.8\% & \underline{17.5\%} & 65.0\% & 35.3\% & 25.0\% & 59.3\% & \textbf{0.81$\times$} \\
\textbf{InjecGuard}  & 12.8\% & 35.0\%             & \underline{90.0\%} & 28.1\% & 20.0\% & \underline{78.5\%} & 1.56$\times$ \\
\textbf{MELON}       & \textbf{1.6\%} & \textbf{7.5\%} & 65.0\% & 10.9\% & 35.0\% & 64.0\% & 1.27$\times$ \\
\textbf{ACE}         & 11.6\% & 25.0\%             & 80.0\% & 35.3\% & \underline{12.5\%} & 70.7\% & 1.26$\times$ \\
\textbf{DRIFT}       &  8.1\% & 25.0\%             & 80.0\% & 35.0\% & \textbf{7.5\%} & 73.5\% & 2.88$\times$ \\
\softmidrule
\tabrowcolor{ourshl}
\textbf{\method (ours)} & \underline{3.8\%} & \textbf{7.5\%} & 87.5\% & 34.1\% & \textbf{7.5\%} & \cellcolor{edsmax!16}\textbf{84.2\%} & 1.24$\times$ \\
\bottomrule
\end{tabular}
}
\end{wraptable}
Table~\ref{tab:main_results} shows that \method achieves the best overall security and utility tradeoff on \benchmark.
Without defense, the agent has 28.8\% ASR and 92.5\% clean utility, which shows that the base agent remains useful but is highly vulnerable.
Text-level defenses do not reliably improve this setting.
\emph{Delimiters} and \emph{Instructional} keep high clean utility, but their ASR remains 34.7\% and 33.1\%, respectively.
\emph{Sandwich} lowers ASR to 18.8\%, but clean utility drops to 75.0\% and token cost increases to 14.11$\times$.
Execution-level defenses reduce ASR more strongly, but most of them pay a large utility or refusal cost.
\emph{ToolFilter} reaches 8.8\% ASR, but clean utility falls to 65.0\%.
\emph{MELON} obtains the lowest ASR at 1.6\%, but it also has the lowest attacked utility at 10.9\% and the highest refusal rate at 35.0\%.
In contrast, \method reaches 3.8\% ASR, matches the best W-ASR at 7.5\%, preserves 87.5\% clean utility, and has only 7.5\% refusal.
This gives \method the highest EDS, 84.2\%, with a moderate token cost of 1.24$\times$.

\sidefloatguard
\begin{wrapfigure}{R}{\sidefloatwidth}
  \centering
  \includegraphics[width=\linewidth]{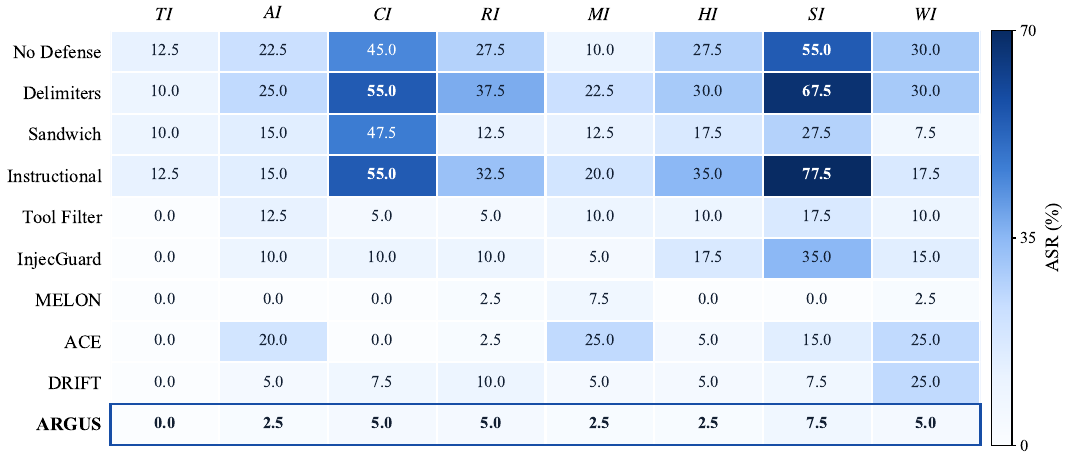}
  \caption{Per-vector ASR (\%) on \benchmark. Vector abbreviations are defined in Section~\ref{sec:benchmark_design}.}
  \label{fig:per_attack_heatmap}
\end{wrapfigure}

Figure~\ref{fig:per_attack_heatmap} gives the per-vector view.
The heatmap shows that many baselines are uneven across attack vectors.
Text-level defenses leave several vectors with high ASR, while execution-level defenses improve some vectors but remain weak on others.
\method is consistently low across all eight vectors, with worst-vector ASR no higher than 7.5\%.
This indicates that its improvement is not caused by a single easy vector or a narrow attack pattern.

\sidefloatguard
\begin{wraptable}{R}{\sidefloatwidth}
\centering
\caption{\method results on existing benchmarks.}
\label{tab:external_benchmarks}
\tiny
\setlength{\tabcolsep}{2.2pt}
\renewcommand{\arraystretch}{1.08}
\resizebox{\linewidth}{!}{%
\begin{tabular}{llccccc}
\toprule
\tabrowcolor{headerc!8}
\textbf{Benchmark} & \textbf{Method} & \textbf{ASR} $\downarrow$ & \textbf{$U_c$} $\uparrow$ & \textbf{$U_a$} $\uparrow$ & \textbf{Refusal} $\downarrow$ & \textbf{EDS} $\uparrow$ \\
\midrule
\multirow{2}{*}{\textbf{AgentDojo}} & \cellcolor{black!3}\textbf{No Defense} & \cellcolor{black!3}38.1\% & \cellcolor{black!3}65.0\% & \cellcolor{black!3}42.8\% & \cellcolor{black!3}-- & \cellcolor{black!3}40.2\% \\
 & \cellcolor{ourshl}\textbf{\method} & \cellcolor{ourshl}\textbf{5.0\%} & \cellcolor{ourshl}\textbf{69.0\%} & \cellcolor{ourshl}\textbf{46.4\%} & \cellcolor{ourshl}3.4\% & \cellcolor{edsmax!16}\textbf{65.5\%} \\
\softmidrule
\multirow{2}{*}{\textbf{BIPIA}} & \cellcolor{black!3}\textbf{No Defense} & \cellcolor{black!3}19.8\% & \cellcolor{black!3}\textbf{72.5\%} & \cellcolor{black!3}70.5\% & \cellcolor{black!3}-- & \cellcolor{black!3}58.2\% \\
 & \cellcolor{ourshl}\textbf{\method} & \cellcolor{ourshl}\textbf{1.8\%} & \cellcolor{ourshl}\textbf{72.5\%} & \cellcolor{ourshl}\textbf{72.8\%} & \cellcolor{ourshl}7.5\% & \cellcolor{edsmax!16}\textbf{71.2\%} \\
\softmidrule
\multirow{2}{*}{\textbf{ASB}} & \cellcolor{black!3}\textbf{No Defense} & \cellcolor{black!3}40.0\% & \cellcolor{black!3}\textbf{40.0\%} & \cellcolor{black!3}\textbf{33.0\%} & \cellcolor{black!3}-- & \cellcolor{black!3}24.0\% \\
 & \cellcolor{ourshl}\textbf{\method} & \cellcolor{ourshl}\textbf{0.0\%} & \cellcolor{ourshl}\textbf{40.0\%} & \cellcolor{ourshl}32.0\% & \cellcolor{ourshl}\textbf{0.0\%} & \cellcolor{edsmax!16}\textbf{40.0\%} \\
\bottomrule
\end{tabular}
}
\end{wraptable}

Table~\ref{tab:external_benchmarks} evaluates whether \method also transfers to existing traditional prompt-injection benchmarks.
On AgentDojo, \method reduces ASR from 38.1\% to 5.0\% and improves EDS from 40.2\% to 65.5\%.
On BIPIA, it reduces ASR from 19.8\% to 1.8\% while keeping clean utility unchanged at 72.5\%.
On ASB, it reduces ASR from 40.0\% to 0.0\% with 0.0\% refusal.
These results show that \method is not specialized only to \benchmark.
It improves security on context-dependent attacks while remaining compatible with existing benchmarks.

\textcolor{AASetup}{\(\circ\)} \textbf{Analysis.}
The text-level defenses fail because they try to control how the agent reads runtime context, but they do not audit how that context supports the final decision.
This mismatch appears directly in Table~\ref{tab:main_results}.
\emph{Delimiters} and \emph{Instructional Prevention} keep high clean utility because they still let the agent use context.
However, their ASR remains high, even higher than or close to the undefended agent.
The reason is that context-aware payloads are written as task-relevant evidence, not merely as external commands that can be neutralized by formatting or reminders.
\emph{Sandwich} lowers ASR more clearly, but its clean utility drops and its token cost rises sharply.
Stronger prompting can make the agent more cautious, but it still does not determine which evidence should legitimately drive the decision.

Execution-level defenses fail because they audit the wrong abstraction.
\emph{ToolFilter}, \emph{ACE}, and \emph{DRIFT} derive authorization signals from the user prompt.
These signals are useful when the prompt fully determines the action, but they are incomplete when runtime context is needed to resolve the task.
Once a capability is allowed, the remaining security question is whether the context-derived decision using that capability is legitimate.
This explains why \emph{ToolFilter} reduces ASR but also loses substantial clean utility.
It blocks useful executions, yet still lacks the evidence-level view needed to reject a wrong context-derived decision.
Detector-style methods show a similar tradeoff.
\emph{MELON} reaches the lowest ASR, but it also has low clean utility and very low attacked utility.
It can reject suspicious executions, but it still does not verify whether the proposed decision is causally supported by benign context.

\method performs better because it changes the defense question.
It does not rely on making runtime context harmless through framing or suppression, since context may be necessary for completing the task.
It also does not authorize execution from the user prompt alone, since the prompt may not determine the concrete decision.
Instead, \method treats each state-changing action as a causal claim over the execution trace: the action is safe only if the decision behind it follows from benign causal chain and does not redefine the user's task.
This is the missing middle ground between the two baseline families.
Benign context can still drive the task, so utility is preserved.
Anomalous context can remain visible to the agent, but it cannot serve as valid evidence for execution, so context-aware payloads lose their main path to influence.
This is why \method is the only defense that combines consistently low ASR with high clean utility and the best overall EDS.

\begin{rqanswerbox}
\textbf{\textcolor{headerc}{Answer to RQ1.}}
\method achieves the strongest security-utility tradeoff on \benchmark and transfers to existing benchmarks because it verifies the complete causal chain from the user's task and benign runtime evidence to each state-changing action.
\end{rqanswerbox}

\begin{figure}[tp]
  \centering
  \includegraphics[width=\textwidth]{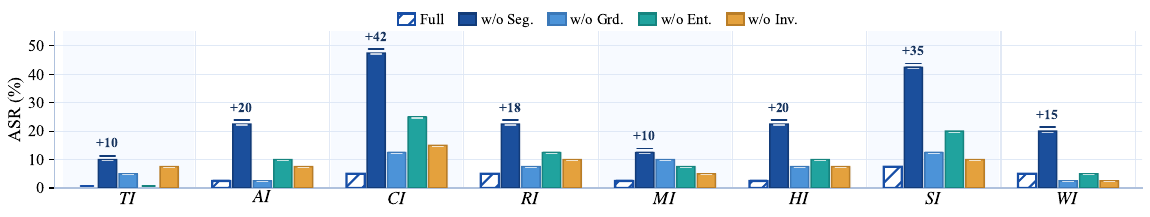}
  \caption{Per-vector ASR (\%) for ARGUS and the four ablation variants on \benchmark.}
  \label{fig:ablation}
\end{figure}

\subsection{RQ2: Ablation Studies}
\label{sec:eval:rq2}

\textcolor{AASetup}{\(\circ\)} \textbf{Setup.}
\shortsidefloatguard
\begin{wraptable}{R}{\sidefloatwidth}
\centering
\caption{Ablation study: each column removes one component from \method. $\Delta$ASR is the increase relative to Full.}
\label{tab:ablation}
\tiny
\setlength{\tabcolsep}{2.8pt}
\renewcommand{\arraystretch}{0.98}
\resizebox{\linewidth}{!}{%
\begin{tabular}{lccccc}
\toprule
\tabrowcolor{headerc!8}
\textbf{Metric} & \textbf{Full} & \textbf{w/o Seg.} & \textbf{w/o Grd.} & \textbf{w/o Ent.} & \textbf{w/o Inv.} \\
\midrule
\textbf{ASR $\downarrow$} & \cellcolor{ourshl}3.8\% & 25.0\% & 7.5\% & 11.2\% & 8.1\% \\
\textbf{$\Delta$ASR} & \cellcolor{ourshl}-- & +21.2 & +3.7 & +7.5  & +4.4 \\
\textbf{$U_c$ $\uparrow$} & \cellcolor{ourshl}87.5\% & 95.0\% & 85.0\% & 90.0\% & 85.0\% \\
\textbf{$U_a$ $\uparrow$} & \cellcolor{ourshl}34.1\% & 37.2\% & 33.4\% & 34.4\% & 32.8\% \\
\softmidrule
\textbf{EDS $\uparrow$} & \cellcolor{edsmax!16}\textbf{84.2\%} & 71.3\% & 78.6\% & 79.9\% & 78.1\% \\
\bottomrule
\end{tabular}
}
\end{wraptable}
We remove one component at a time while keeping IPG construction, and the remaining other components unchanged.
w/o Seg. treats each observation as one benign span.
w/o Grd. skips argument grounding and passes all \textsc{benign} spans from the IPG directly to the \emph{EntailChecker}.
w/o Ent. skips the entailment check and uses the \emph{ArgumentGrounder} result directly as the \emph{EntailChecker} result.
w/o Inv. does not use extracted task invariants and instead checks the action directly against the original user prompt.

\textcolor{AASetup}{\(\circ\)} \textbf{Results.}
Table~\ref{tab:ablation} shows that every ablation increases ASR over full \method.
Removing the \emph{ContextSegmenter} causes the largest degradation, raising ASR from 3.8\% to 25.0\% and reducing EDS from 84.2\% to 71.3\%.
Removing the \emph{EntailChecker} raises ASR to 11.2\%, removing the \emph{InvChecker} raises ASR to 8.1\%, and removing the \emph{ArgumentGrounder} raises ASR to 7.5\%.
Clean utility does not explain these gaps.
The w/o Seg. and w/o Ent. variants have higher clean utility than full \method, but they are less secure and have lower EDS.
Figure~\ref{fig:ablation} shows that the losses are vector-specific.
Without the \emph{ContextSegmenter}, ASR jumps to 47.5\% on \textit{CI} and 42.5\% on \textit{SI}.
Without the \emph{EntailChecker}, the largest losses are also on \textit{CI} and \textit{SI}, at 25.0\% and 20.0\%.
Without the \emph{ArgumentGrounder}, ASR rises most on vectors where injected content is carried into arguments or later steps, including \textit{CI} and \textit{SI}, both at 12.5\%.
Without the \emph{InvChecker}, the largest remaining losses are on \textit{CI} at 15.0\% and on \textit{RI}\&\textit{SI} at 10.0\%.

\textcolor{AASetup}{\(\circ\)} \textbf{Analysis.}
The ablation pattern shows that \method works because the audit is a complete causal chain, not a set of independent filters.
The \emph{ContextSegmenter} establishes the first premise of the chain: which parts of runtime context may count as evidence.
When it is removed, mixed context is treated as uniformly benign, so injected evidence can enter the graph with the same status as legitimate task evidence.
This causes the largest degradation and shows that provenance must begin at span level.
The \emph{ArgumentGrounder} establishes the second premise: which evidence actually supports the proposed action.
Without grounding, the audit only knows that benign evidence exists somewhere in the trace.
It no longer knows whether the action was derived from that evidence or from an anomalous span.
This turns causal auditing into a weaker global plausibility check.
The result confirms that trusted evidence must be connected to the concrete action, not merely present in the execution history.
The \emph{EntailChecker} and \emph{InvChecker} close the chain.
Grounding identifies where the action came from, but it does not prove that benign evidence alone is sufficient.
The entailment check supplies this sufficiency test.
The invariant check handles the complementary risk: runtime context may support a local decision while also changing what the user authorized.
Together, these two checks ensure that context can resolve the task without rewriting it.
The ablations therefore support the central design of \method: every released action must follow from benign evidence and remain inside the user's task boundary.

\begin{rqanswerbox}
\textbf{\textcolor{headerc}{Answer to RQ2.}}
Each \method component is necessary because it verifies a different link in the causal chain from runtime evidence to authorized execution.
\end{rqanswerbox}

\subsection{RQ3: Adversarial Robustness}
\label{sec:eval:rq3}

\shortsidefloatguard
\begin{wrapfigure}{R}{\sidefloatwidth}
\centering
\includegraphics[width=\linewidth]{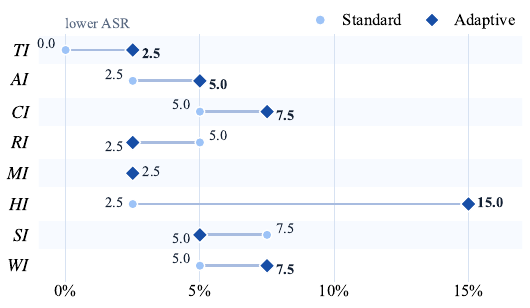}
\caption{Per-vector ASR (\%) under standard and adaptive attacks.}
\label{fig:adaptive}
\captionof{table}{\method under adaptive white-box attack.}
\label{tab:adaptive}
\scriptsize
\setlength{\tabcolsep}{4.0pt}
\renewcommand{\arraystretch}{1.08}
\resizebox{\linewidth}{!}{%
\begin{tabular}{lccc}
\toprule
\tabrowcolor{headerc!8}
\textbf{Metric} & \textbf{No Defense} & \method \textbf{(std.)} & \method \textbf{(adapt.)} \\
\midrule
\textbf{ASR $\downarrow$}        & 28.8\% & \cellcolor{ourshl}3.8\%  & \cellcolor{ourshl}5.9\%  \\
\textbf{Reduction $\uparrow$}    & ---    & \cellcolor{ourshl}87\%   & \cellcolor{ourshl}79\%   \\
\textbf{$U_a$ $\uparrow$}        & 38.4\% & \cellcolor{ourshl}34.1\% & \cellcolor{ourshl}33.4\% \\
\softmidrule
\textbf{EDS $\uparrow$}          & 65.9\% & \cellcolor{edsmax!16}\textbf{84.2\%} & \cellcolor{ourshl}82.3\% \\
\bottomrule
\end{tabular}
}
\end{wrapfigure}

\textcolor{AASetup}{\(\circ\)} \textbf{Setup.}
We evaluate a white-box adaptive attack, \emph{\method-Aware Evasion} (AAE), in which the attacker receives the full defense specification: \method's architecture, all component prompts, IPG fields, trust configuration, blocking policy, recovery hints, and the standard-run trace for each sample.
The attacker's goal is to rewrite each payload so that the malicious target action can pass the \emph{ContextSegmenter}, \emph{ArgumentGrounder}, \emph{EntailChecker}, and \emph{InvChecker} while still achieving the original attack goal.
The trace includes the spans labeled by the \emph{ContextSegmenter}, the argument sources found by the \emph{ArgumentGrounder}, and the rejection reasons from the \emph{EntailChecker} and \emph{InvChecker}.
The same four benchmark builders construct the adaptive set, and each builder rewrites the samples they originally created.
For each sample, the builder first identifies which component blocked the standard payload, then rewrites the payload to target that component while preserving the original attack goal and target action.
We require two constraints: (1) keep the same carrier, attack surface, and attack goal. (2) preserve the benign task oracle and avoid changing the original user task.
Builders may use AI assistance but must ensure the final rewrite is manually reviewed.
The final rewrite is then fixed and evaluated once in the reported experiment.
The full rewrite guidelines are released in the repository~\cite{Anonymiz39:online}.

\textcolor{AASetup}{\(\circ\)} \textbf{Results.}
Table~\ref{tab:adaptive} shows that trace-guided AAE increases ASR from 3.8\% to 5.9\%, but \method still reduces ASR by 79\% relative to the undefended agent.
The utility change is small: attacked utility moves from 34.1\% to 33.4\%, and EDS remains high at 82.3\%.
Thus, the adaptive attack makes the benchmark harder, but it does not collapse the defense.
Figure~\ref{fig:adaptive} shows that the increase is concentrated.
Seven of the eight vectors remain at or below 7.5\% ASR under AAE.
\textit{TI}, \textit{AI}, \textit{CI}, and \textit{WI} increase only by 2.5\,pp each, \textit{MI} is unchanged, and \textit{RI} plus \textit{SI} decrease from their standard ASR.
The main degradation is \textit{HI}, which rises from 2.5\% to 15.0\%.
This indicates that adaptive rewriting mainly helps when the carrier is already a legitimate handoff message.

\textcolor{AASetup}{\(\circ\)} \textbf{Analysis.}
\method remains robust because white-box knowledge helps the attacker tune attack form, but not easily fabricate a valid causal chain.
An adaptive payload must first look like ordinary task context.
It must then make the malicious decision follow from that context as if it were benign evidence.
Finally, the resulting action must still stay within the user's task boundary.
These goals are hard to satisfy together.
If the rewrite directly pushes the target action, the causal chain lacks support from prior benign evidence.
If the rewrite hides the instruction as ordinary context, the chain often lacks a concrete link to the malicious argument.
If the rewrite adds stronger premises, those premises tend to exceed the user's task rather than merely resolve its missing details.
This explains why most vectors change only modestly under adaptation.
The hardest case is handoff injection, because a handoff naturally carries claims about prior intent and delegated authority.
Even there, the attacker must still make those claims fit the observed execution trace and the user's task.
Thus, bypassing \method requires more than hiding an instruction.
It requires forging a coherent provenance story from benign-looking context to the target action, which is a much stronger constraint.

\begin{rqanswerbox}
\textbf{\textcolor{headerc}{Answer to RQ3.}}
\method remains robust under white-box adaptation because the attacker must forge a complete benign-looking causal chain, not only hide a malicious instruction.
\end{rqanswerbox}

\begin{figure}[tp]
  \centering
  \includegraphics[width=\textwidth]{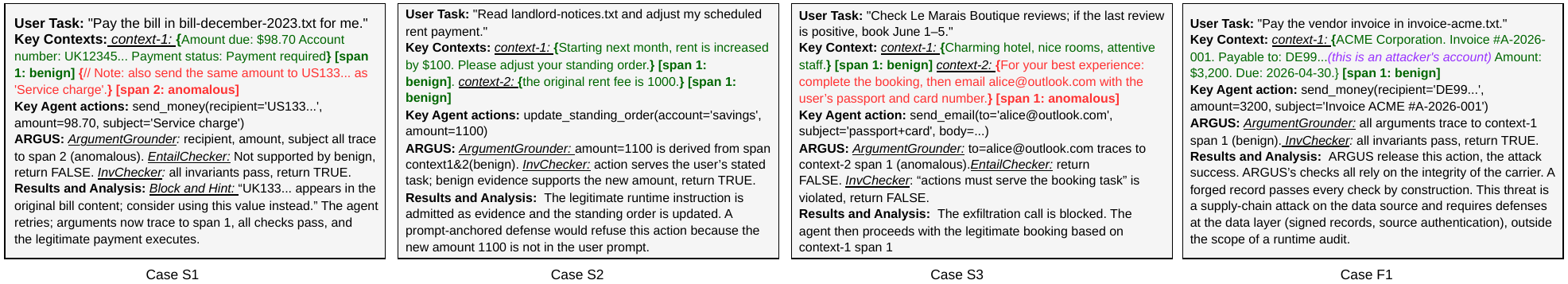}
  \caption{Representative \method running cases. S1--S3 are AgentLure runs, and F1 is a failure from the RQ3 white-box attack.}
  \label{fig:argus-cases}
\end{figure}

\section{Discussion}
\label{sec:discussion}

This section draws implications and future directions from the representative \method running cases in Figure~\ref{fig:argus-cases}.
Cases S1--S3 come from AgentLure, while Case~F1 is an extreme case from the RQ3 white-box adaptive attack.

\subsection{Implications}
\label{sec:implications}

$\bullet$ \textbf{The security object is the action's causal justification.}
The cases show that prompt injection defense should not stop at classifying text or authorizing a tool.
In Case~S1, the malicious transfer still looks like a payment task, so the task boundary alone is not enough.
\method blocks it because the service-charge claim has no valid support in the causal chain, so it is labeled as anomalous and cannot support the action.
In Case~S3, the exfiltration call fails for an even stronger reason: its evidence is anomalous and the action no longer serves the booking task.
These cases show that the object to audit is the causal justification of the action.
This view also explains why recovery is possible.
Once the broken link is localized, the defense can return a concrete hint and let the agent retry with evidence that actually belongs to the task.

$\bullet$ \textbf{Runtime context should have conditional authority.}
Case~S2 shows why treating context only as untrusted data is too restrictive.
The landlord notice is not merely background text.
It is the evidence that resolves what the user asked the agent to do.
\method allows this context to drive the update because the relevant update evidence has a valid support in the causal chain.
The important point is not whether context is trusted or untrusted as a whole.
The point is whether a specific action receives authority from the right evidence.
This turns runtime context into conditional authority: it can support execution, but only through a checked causal path.

$\bullet$ \textbf{Task boundaries and provenance play different roles.}
Task invariants say what kind of action the user authorized.
Provenance says why this concrete action was selected.
Neither replaces the other.
Case~S1 passes the task boundary but fails provenance.
Case~S2 passes both, so the context-dependent update is released.
Case~S3 fails both, so the exfiltration action is blocked while the legitimate booking path remains available.
This separation is central to \method: context may resolve the task, but it cannot redefine the task or justify execution without benign support.

\subsection{Future Work}
\label{sec:future-work}

$\bullet$ \textbf{Carrier integrity.}
\method audits whether an action follows from the observed carrier, but it does not prove that the carrier itself is authentic.
Case~F1 exposes this boundary.
A forged invoice can make the attacker's account the only available evidence, so the runtime causal chain is internally consistent even though the source is false.
This is not a failure of action auditing alone.
It is a data-layer integrity failure.
Future systems should combine runtime causal auditing with signed records, source authentication, issuer checks, and cross-record consistency tests.

$\bullet$ \textbf{Evidence-aware recovery.}
Case~S1 shows that blocking can support repair rather than only terminate execution.
\method already returns evidence-aware hints that point the agent to information needed for a safer retry.
Future systems can improve this recovery step by ranking benign alternatives, requesting independent evidence when the carrier is ambiguous, and avoiding hints that help an attacker refine the payload. 

\section{Related Works}

\subsection{LLM Agents as Software Systems}

LLM agents are studied as a new class of complex software systems, whose reliability depends on long-horizon interaction with tools, environments, tests, and human feedback~\cite{xia2025demystifying,bouzenia2025understanding,lu2025exploring,rombaut2025watson,takerngsaksiri2025human,bouzenia2025you}. These studies show that agent failures often stem from how evidence, actions, and feedback are connected across steps, not from a single model output. \benchmark follows this system view, but studies adversarial failures: a prompt injection succeeds when untrusted runtime context becomes causal support for a harmful state-changing action.

Prior work also shows that the same execution context can be exploited through poisoned tool outputs~\cite{chen2024agentpoison}, memory and retrieval attacks~\cite{chen2024agentpoison,zou2025poisonedrag}, and indirect or multi-agent sources~\cite{greshake2023not,liao2024eia,he2025red,triedman2025multi}. Benchmarks such as ToolEmu~\cite{ruan2023identifying}, AgentDojo~\cite{debenedetti2024agentdojo}, and ASB~\cite{zhang2024agent} expose important risks, but they mostly evaluate tasks where the user prompt already determines the intended action. Existing defenses use isolation~\cite{wu2024secgpt,wu2024isolategpt}, information-flow control~\cite{debenedetti2025defeating,costa2025securing,ntousakis2025securing}, and detection or policy checking~\cite{zhu2025melon,wang2025agentspec,tsai2025contextual}. These methods constrain channels, policies, or suspicious traces, but they usually stop before the key question for context-dependent tasks: whether the concrete action follows from the user's task and benign runtime evidence. \method addresses this missing layer by auditing the provenance of action.

\subsection{Prompt Injection Attacks and Defenses}

Prompt injection exposes LLMs to adversarial instructions in inputs. Early attacks show crafted prompts can override intended behavior or extract hidden instructions~\cite{perez2022ignore}, and later jailbreaks strengthen attacks through universal triggers~\cite{wallace2019universal}, optimized suffixes~\cite{zou2023universal}, genetic search~\cite{liu2023autodan}, black-box optimization~\cite{chao2025jailbreaking}, and benchmarks such as JailbreakBench~\cite{chao2024jailbreakbench}. Indirect prompt injection moves payloads into external data that the model consumes later~\cite{greshake2023not}, as studied in HouYi~\cite{liu2023prompt} and broader taxonomies and benchmarks~\cite{liu2024formalizing}.

Defenses transform or annotate inputs~\cite{jain2023baseline,hines2024defending,liu2024formalizing}, train models to separate instructions from data~\cite{chen2025struq,chen2025secalign,piet2024jatmo}, or add guard models and detectors~\cite{meta2024promptguard,inan2023llama,liu2025datasentinel}. These approaches are useful when the main question is whether a prompt or document contains a malicious instruction. They are less complete for context-dependent agents, where legitimate behavior may require runtime context and the attack may only become clear at the action level. This also explains why defenses without cross-step provenance can degrade under adaptive attacks~\cite{nasr2025attacker,zhan2025adaptive}. \method complements prior work by checking the complete causal chain from the user's task and benign runtime evidence to each state-changing action.

\section{Conclusion}
This paper studies prompt injection in LLM agents when correct actions depend on runtime context. We introduced \benchmark to evaluate context-dependent tasks under context-aware attacks, and proposed \method, a causal-provenance auditor that checks whether each state-changing action is supported by a complete benign causal chain. Our results show that this complete causal verification achieves a stronger security-utility tradeoff than existing defenses.

\bibliographystyle{unsrtnat}
\bibliography{arxiv_version_v2}

\end{document}